\documentclass[showpacs, twocolumn, prl, floatfix]{revtex4}

\usepackage{amssymb,amsmath} 
\usepackage{stmaryrd}
\usepackage{bbm} 
\usepackage{bm} 
\usepackage[english]{babel}
\usepackage{slashed}

\usepackage{graphicx}
\usepackage{color}

\usepackage{float}

\usepackage[bookmarks,breaklinks,plainpages=false,pdfpagelabels]{hyperref}  
	\hypersetup{linkcolor=red,citecolor=blue,filecolor=dullmagenta,urlcolor=darkblue} 

\definecolor{grey}{rgb}{0.4,0.4,0.4}
\definecolor{dullmagenta}{rgb}{0.4,0,0.4}
\definecolor{darkblue}{rgb}{0,0,0.4}
\definecolor{orange}{rgb}{1,0.5,0}
\definecolor{lightbrown}{rgb}{0.75,0.5,0.25}
\definecolor{tan}{cmyk}{0.14,0.42,0.56,0}
\definecolor{djunglegreen}{cmyk}{0.99,0,0.52,0}
\definecolor{lightgreen}{rgb}{0,1,0}
\definecolor{olivegreen}{cmyk}{0.64,0,0.95,0.40}
\definecolor{midgreen}{rgb}{0.0,0.675,0.0}

\newcommand{\normsingle}[1]{\left|{#1}\right|}

\newcommand{\e}{\ensuremath{\epsilon}}

\renewcommand{\l}{\ensuremath{\lambda}}
\newcommand{\s}{\ensuremath{\sigma}}

\newcommand{\q}{\quad}

\newcommand{\qqq}{\qquad\quad}

\newcommand{\vs}{\vspace}

\renewcommand{\.}{\hspace{0.5mm}}

\newcommand{\ra}{\ensuremath{\rightarrow}}

\newcommand{\Frm}{\ensuremath{\mathrm{F}}}
\newcommand{\Grm}{\ensuremath{\mathrm{G}}}

\newcommand{\arm}{\ensuremath{\mathrm{a}}}

\newcommand{\hrm}{\ensuremath{\mathrm{h}}}
\newcommand{\irm}{\ensuremath{\mathrm{i}}}
\newcommand{\jrm}{\ensuremath{\mathrm{j}}}

\newcommand{\Lcal}{\ensuremath{\mathcal{L}}}

\newcommand{\Ocal}{\ensuremath{\mathcal{O}}}
\newcommand{\Pcal}{\ensuremath{\mathcal{P}}}

\newcommand{\Scal}{\ensuremath{\mathcal{S}}}

\newcommand{\Zcal}{\ensuremath{\mathcal{Z}}}

\newcommand{\Jbbm}{\ensuremath{\mathbbm{J}}}

\newcommand{\kbm}{\ensuremath{\bm{k}}}

\newcommand{\xbm}{\ensuremath{\bm{x}}}

\newcommand{\D}{\ensuremath{\mathcal{D}}}
\renewcommand{\d}{\ensuremath{\mathrm{d}}}
\newcommand{\ee}{\ensuremath{\mathrm{e}}}

\newcommand{\defas}{\mathrel{\mathop :}=} 

\newcommand{\Tr}{\ensuremath{\mathrm{Tr}}}
\newcommand{\diag}{\ensuremath{\mathrm{diag}}}

\renewcommand{\t}{\text}

\newcommand{\eg}{e.g.}

\newcommand{\hc}{{\rm h.c.}}

\newcommand{\cf}{c.f.}

\begin{document}

\title{Large-Scale Suppression from Stochastic Inflation}

\author{Florian K{\"u}hnel}
\email{kuehnel@physik.uni-bielefeld.de}
\author{Dominik J.~Schwarz}
\email{dschwarz@physik.uni-bielefeld.de}
\affiliation{Fakult{\"a}t f{\"u}r Physik, Universit{\"a}t Bielefeld, Postfach 100131, 33501 Bielefeld, Germany}
\date{\today}

\begin{abstract}
We show non-perturbatively that the power spectrum of a self-interacting scalar field in de Sitter space-time is strongly suppressed on large scales. The cut-off scale depends on the strength of the self-coupling, the number of $\ee$-folds of quasi-de Sitter evolution, and its expansion rate. As a consequence, the two-point correlation function of field fluctuations is free from infra-red divergencies.
\end{abstract}


\maketitle

A central building block of our current understanding of the Universe is the idea of cosmological inflation. It can be realized by a single scalar field, called inflaton. The quantum fluctuations of this field and of the space-time seed today's structures in the Universe. This is well understood in the case of small fluctuations. However, some models of inflation imply the existence of large quantum fluctuations, \eg, the so-called chaotic scenario \cite{Mod.Phys.Lett.A1.81.5}, which also leads to the idea of eternal inflation \cite{PhysRevD.27.2848}.

Our task is to study the implications of large quantum fluctuations on structure formation and the evolution of the Universe. Of special importance is the power spectrum, which is observable via the cosmic microwave background radiation \cite{Spergel:2003cb}.

A toy model to understand the physics of large quantum fluctuations is a self-interacting scalar field $\Phi$ in de Sitter space-time. In the traditional approach, based on perturbative quantum field theory in curved space-time, the multi-point correlation functions of $\Phi$ generically exhibit infra-red divergencies (\cf~\cite{Lyth:2007jh}).

Related to these issues is the presence of large fluctuations over super-horizon distances, which can invalidate the entire semi-classical approximation \cite{Burgess:2010dd}. This breakdown of semic-classical methods is due to the fact that higher loops are not suppressed, meaning that all orders in the loop expansion have comparable sizes. Hence, non-peturbative methods are needed for a full description and deeper understanding of inflationary cosmology.

Stochastic inflation \cite{1986LNP...246..107S, Goncharov:1987ir, 1987NuPhB.284..706R, Starobinsky:1994bd, linde-1994-49, Woodard:2005cw, martin-2006-73-a, kuhnel-2008, kuhnel-2009} provides one of the very few of such approaches. Its idea lies in splitting the quantum fields into long- and short-wavelength modes, and viewing the former as classical objects evolving stochastically in an environment provided by quantum fluctuations of shorter wavelengths. Given the de Sitter horizon $c / H$ as a natural length scale of the problem, one then focusses on the ``relevant'' degrees of freedom (the long-wavelength modes) and regards the short-wavelength modes as ``irrelevant'' ones, where ``short'' and ``long'' are subject to the horizon.

The most simple setup provides a fixed cosmological background, in which the dynamics of a scalar test field $\Phi$ is analyzed. If $\Phi$ is free, massive and minimally coupled, one obtains after splitting into long and short wavelengths, $\Phi = \varphi + \phi$, an effective equation of motion of generalized Langevin-type,
\begin{align}
	\left( \Box + \mu^{2} \right) \varphi
		&=						\hrm
								.
								\label{eq:linearfieldeq}
\end{align}
The quantity $\hrm$ is a Gaussian-distributed random force with zero mean.

In \cite{kuhnel-2008, kuhnel-2009} we applied replica field theory together with a Gaussian variational method to stochastic inflation. Extending early studies, that mainly focussed on homogeneous fields and thus restricting attention to the time evolution of $\Phi$, we presented a method to calculate arbitrary two-point correlation functions.

In this work we extend our previous results \cite{kuhnel-2008, kuhnel-2009} to include self-interactions of a scalar field in de Sitter space-time. For the specific example of a quartic self-interaction we calculate the power spectrum and show that self-couplings cause a damping of this quantity on large scales. This therefore solves the problem of infra-red divergencies of two-point correlation functions. Furthermore, our results are independent of the precise way of splitting into long- and short-wavelength modes and hence do not suffer from any related ambiguity.

As a starting point, we use the Lagrangian
\begin{align}
	\Lcal
		&=						\frac{1}{2}\,g^{\mu\nu}\partial_{\mu}\Phi\.\partial_{\nu}\Phi
								- \lambda\.\Phi^{4}
								,
								\label{eq:L}
\end{align}
where $\Phi$ is a massless, minimally-coupled, real scalar field with quartic self-coupling constant $\lambda$. Greek indices run from $0$ to $3$. We assume a de Sitter background geometry, $( g_{\mu\nu} ) = \diag( 1, - a( t )^{2}, - a( t )^{2}, - a( t )^{2} )$ with the scale factor $a( t ) = \exp( H\mspace{1mu}t )$. For convenience we use $\hslash = c = 1$.

Let $\Phi^{}_{0}$ be a free field, being subject to \eqref{eq:L} with $\lambda = 0$. It might be decomposed as
\begin{align}
	\Phi^{}_{0}( t, \kbm )
		&=						\hat{\arm}( \kbm )\. u^{}_{0}( t, k ) + \hc
								,
								\label{eq:Phi-fourier-decomposition}
\end{align}
with the modulus of the comoving momentum $k \defas \normsingle{\kbm}$. The annihilation and creation operators, $\hat{\arm}$ and $\hat{\arm}^{\dagger}$, obey the usual commutation relations.

In terms of conformal time $\tau$, with $a( \tau ) = ( H\mspace{1mu}\tau )^{-1}$, the rescaled mode functions $v^{}_{0}( \tau, k ) \defas a( \tau )\. u^{}_{0}( \tau, k )$ fulfil the mode equation
\begin{align}
	v_{0}'' + \!\left[ k^{2} - \frac{2}{\tau^{2}} \right]\! v^{}_{0}
		&=						0
								,
								\label{eq:modeequation-v0}
\end{align}
where primes denote derivatives with respect to $\tau$. Solutions to \eqref{eq:modeequation-v0} are fixed by requiring that for very short wavelengths the effect of space-time curvature becomes irrelevant, and thus a plane-wave solution should be obtained, $\lim_{k / a \,\ra\, \infty} v^{}_{0}( \tau, k ) = \ee^{\irm k \tau}/\sqrt{2 k}$. The factor $1 / \sqrt{2\,k}$ is fixed by the canonical commutation relations of $\Phi^{}_{0}$ and its conjugate momentum. At late times, the leading term of the solution to \eqref{eq:modeequation-v0} reads
\begin{align}
	u^{}_{0}\big( k \ll 1/ | \tau | \big)
		&\simeq					-\; \frac{ \irm\.H }{ \sqrt{2\.k^{3}\,} }
								.
								\label{eq:u0-late-time}
\end{align}

An object of central interest in cosmology is the power spectrum $\Pcal( k )$. Its relation to the propagator
\begin{align}
	\Grm( k )\.( 2 \pi )^{3}\.\delta^{3}\!\left( \kbm - \kbm' \right)
		&\equiv					\big\langle \Omega \big| \Phi( \kbm)\.
								\Phi( \kbm' ) \big| \Omega \big\rangle
								,
\end{align}
where $\big| \Omega \big\rangle$ is the Bunch-Davies vacuum, is given by
\begin{align}
	\Pcal( k )
		&\defas					\frac{ k^{3} }{ 2 \pi^{2} }\. \Grm( k )
								.
								\label{eq:power-spectrum}
\end{align}
On superhorizon scales $\big( k \ll 1/ | \tau | \big)$ one finds for the free massless case (subscript ``0'') a scale-invariant spectrum:
\begin{align}
	\Pcal^{}_{0}( k )
		&=						\frac{ k^{3} }{ 2 \pi^{2} } \big| u^{}_{0}( k ) \big|^{2}
		=						\frac{ H^{2} }{ ( 2 \pi )^{2} }
								.
								\label{eq:power-spectrum-free-case}
\end{align}

To go beyond Equation \eqref{eq:power-spectrum-free-case}, we first split the field $\Phi$ into a short- and a long-wavelength part, $\Phi = \phi + \varphi$, where we use the filter function $\Frm_{\!\kappa}$, specified by its derivative:
\begin{align}
	\Frm_{\!\kappa}'( y )
		&=						\begin{cases}
									0						& : \. y < - \kappa ,\\
									N \exp\!\left( 1 - \left[ 1 - \left( \frac{ y }{ \kappa } \right)^{\!2} \right]^{\!-2} \right)
															& : \. y \in [ -\kappa, \kappa ] ,\\
									0						& : \. y > + \kappa ,
								\end{cases}
								\label{eq:filter-function}
\end{align}
with $y \defas k | \tau | - \e$, cutting out wave numbers below $\e / | \tau |$ with a cutting width $\kappa$. In the limit $\kappa \ra 0$, $\Frm_{\!\kappa}$ approaches the step function $\delimiterfactor=1010 \Theta\!\left( k | \tau | - \e \right)$. The constant $N \defas \ee\.\sqrt{\pi\,} \sqrt[4\;]{2\,} / ( 5.3\.\kappa )$ in \eqref{eq:filter-function} is a normalization factor, ensuring $\int\!\d y\,\Frm_{\!\kappa}'( y ) = 1$. Throughout this work we choose $\kappa = 10^{-3}$ and $\e = 10^{-2}$, although our main statements are virtually independent of these quantities.

Having introduced the precise way of splitting into long- and short-wavelength modes with the filter function \eqref{eq:filter-function}, we now consider the form of the induced noise terms. For general self-interactions, they are non-linear in the short-wavelength modes. However, if one is interested in the late-time behavior, or more precisely in the leading-$\ln\!\big( a( t ) \big)$ contribution, one may restrict to linear, Gaussian-distributed noise terms. This has been argued already a long time ago by Starobinsky \cite{1986LNP...246..107S} and has been rigorously proven by Woodard \cite{Woodard:2005cw}.

The stochastic field equation for model \eqref{eq:L} reads then
\begin{align}
	\Box \varphi + 4\,\lambda\,\varphi^{3}
		&=						\hrm
								,
								\label{eq:cubicfieldeq}
\end{align}
where $\hrm$ is a Gaussian-distributed random variable with
\begin{align}\label{eq:linear-distribution}
	\overline{\hrm}
		&=						0,\q
	\overline{\hrm^{2}}
		=						\Delta
								,
\end{align}
where $\Delta$ is a known function, depending on derivatives of the mode functions in \eqref{eq:Phi-fourier-decomposition}, see \eg, \cite{1987NuPhB.284..706R} for a more detailed presentation. The ``bar'' in \eqref{eq:linear-distribution} and below denotes the average over the noise due to quantum fluctuations of short wavelengths.

Different filter functions have been intensively discussed in \cite{kuhnel-2009}. In \cite{kuhnel-2008} we showed for free fields that filter functions with compact support allow us to avoid infra-red divergencies. Below we show that they are also absent if quartic self-interactions are turned on. In fact, this result in independent of the choice of a filter function.

Let us now briefly summarize the program we will perform next: As we described in detail in \cite{kuhnel-2008, kuhnel-2009}, we first Wick-rotate to Euclidean signature and use the replica trick \cite{1988PhT....41l.109M}. Then we introduce a suitable variational action and determine its form (especially its replica structure) from a Feynman-Jensen variational principle \cite{PhysRev.97.660}. This will allow us to go beyond ordinary perturbation theory (\cf~\cite{1991JPhy1...1..809M, kuhnel-2008, kuhnel-2009}) and to obtain an analytic expression for the full power spectrum.

After Wick-rotating we proceed with the replica trick,
\begin{align}
\begin{split}
	&\frac{\delta^{n} }{\delta \jrm( x_{1} ) \ldots \delta \jrm( x_{n} )}\. \overline{ \ln\big( \Zcal[ \mspace{1mu}\jrm\mspace{1mu} ] \big) }\\
		&\qqq=					\lim_{m \rightarrow 0} \frac{1}{m} \frac{\delta^{n} }{\delta \jrm( x_{1} ) \ldots \delta \jrm( x_{n} )}
								\ln\!{\left( \overline{ \Zcal^{m}[ \mspace{1mu}\jrm\mspace{1mu} ] } \right)}
								,
								\label{eq:relica-trick}
\end{split}
\end{align}
where $\Zcal[ \mspace{1mu}\jrm\mspace{1mu} ]$ is the generating functional depending on an external current $\jrm$. $m$ denotes the number of replicas, labelled by the indices $a, b, \ldots$\;\.. Furthermore we define the replicated action ${\Scal}^{(m)}$ via
\begin{align}
\begin{split}
	\overline{\Zcal^{m}[ \mspace{1mu}\jrm\mspace{1mu} ]}
		&=						\int{\prod_{a=1}^{m}\!\D[ \varphi^{}_{a} ]}\, \overline{ \exp\!{ \left( - \sum_{b=1}^{m}
								\Scal\big[ \varphi^{}_{b}, \jrm\mspace{1mu} \big] \right) } }\\
		&\equiv					\int{\prod_{a=1}^{m}\!\D[ \varphi^{}_{a} ]}\,
								\exp\!{ \left(\! - {\Scal}^{(m)}\big[ \{ \varphi \}, \jrm\mspace{1mu} \big] \right) }
								.
								\label{eq:Zmbar}
\end{split}
\end{align}
Besides terms diagonal in replica space ($\propto\!\delta_{ab}$), it also contains the non-diagonal part
\begin{align}
	&{\Scal}^{(m)}\big[ \{ \varphi \}, \jrm\mspace{1mu} \big]
		\supset					- \frac{1}{2}\sum_{a,b = 1}^{m}\int_{t, \kbm}\varphi^{}_{a}( t, \kbm )\,
								{\Delta}( t, \kbm )\,\varphi^{}_{b}( t, -\kbm )
								,
								\label{eq:S(m)}
\end{align}
originating from the average over noise.

We apply the Feynman-Jensen variation principle and therefore define a Gaussian variational action
\begin{equation}
	\Scal^{(m)}_{\t{\tiny var}}\big[ \{ \varphi \} \big]
		\defas					\frac{1}{2}\sum_{a,b = 1}^{m}\int_{t, \kbm}\varphi^{}_{a}( t, \kbm )\,
								{\Grm^{-1}}_{ab}( t, \kbm )\,\varphi^{}_{b}( t, -\kbm )
								,
\end{equation}
with $\int_{t} \defas \int\d t$ and $\int_{\kbm} \defas \int\d^{3} k / ( 2 \pi )^{3}$. We make the ansatz for the inverse propagator
\begin{align}
	{\Grm^{-1}}_{ab}
		&\defas					\Big[ \Grm_{0}^{-1} + \s \Big] \delta_{ab} - \s_{ab}
								.
								\label{eq:Gab-inverse-ansatz}
\end{align}
The self-energy matrix $\llbracket \s\mspace{1mu}\delta_{ab} - \s_{ab} \rrbracket$ mimics the diagonal and the non-diagonal parts in \eqref{eq:Zmbar}, respectively.

Maximizing the right-hand side of the Feynman-Jensen inequality
\begin{align}
	\ln( \Zcal )
		&\ge						\ln( \Zcal^{}_{\t{\tiny var}} )
								+ \Big\langle \Scal^{(m)}_{\t{\tiny var}} - \Scal^{(m)} \Big\rangle_{\!\t{\tiny var}}
								,
								\label{eq:fvar}
\end{align}
wherein the subscript ``var'' refers to the variational action \eqref{eq:S(m)}, yields the replica symmetric solution
\begin{subequations}
\begin{align}
	\s
		&\simeq					\frac{ 6\.\lambda\.H }{ m } \int_{\kbm} \Tr\llbracket \Grm_{ab} \rrbracket
								,
								\label{eq:sc=-Delta+4lambdaGaa}\\
	\s_{ab}
		&\simeq					\Delta\.\Jbbm_{ab}
								,
								\label{eq:s=sabDelta}
\end{align}
\end{subequations}
with $a \ne b$. The $m \times m$-matrix $\Jbbm$ is defined by $\Jbbm_{ab} = 1$ for all $a, b$.

To solve the implicit equations \eqref{eq:sc=-Delta+4lambdaGaa} and \eqref{eq:s=sabDelta} we invert \eqref{eq:Gab-inverse-ansatz} by means of an expansion in the number of replicas $m$. At leading order we find
\begin{align}
	\Grm_{ab}
		&\simeq					\Big[ \Grm_{0}^{-1} + \s \Big]^{-1} \delta_{ab}
								+ \Delta\.\Big[ \Grm_{0}^{-1} + \s \Big]^{-2} \Jbbm_{ab}
								.
								\label{eq:Gab}
\end{align}
From this quantity we extract the full physical propagator $\Grm( t, k )$ via (\cf~\cite{1991JPhy1...1..809M})
\begin{align}
	\Grm( t, k )
		&=						\lim_{m \ra 0} \frac{ 1 }{ m } \Tr\big\llbracket \Grm_{ab}( t, k ) \big\rrbracket
								.
								\label{eq:Gphys}
\end{align}

\begin{figure}
	\centering
	\includegraphics[angle=0,scale=1]{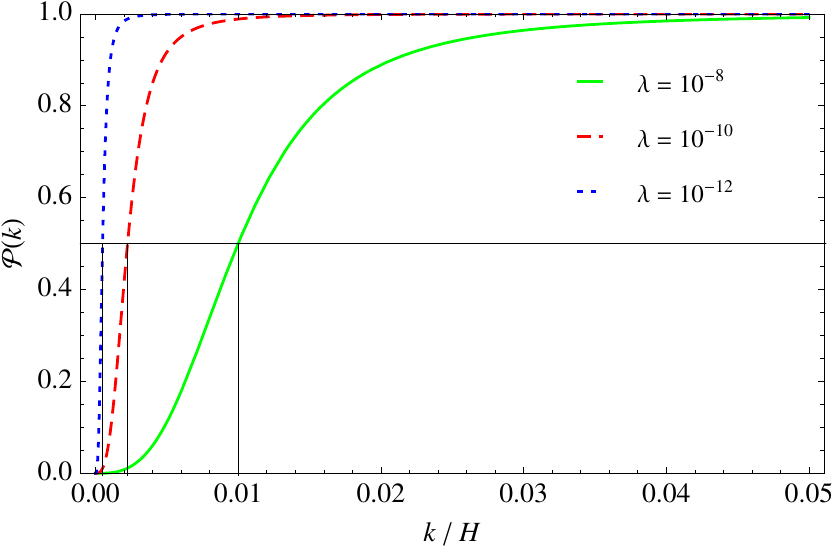}\\[2mm]
	\includegraphics[angle=0,scale=1]{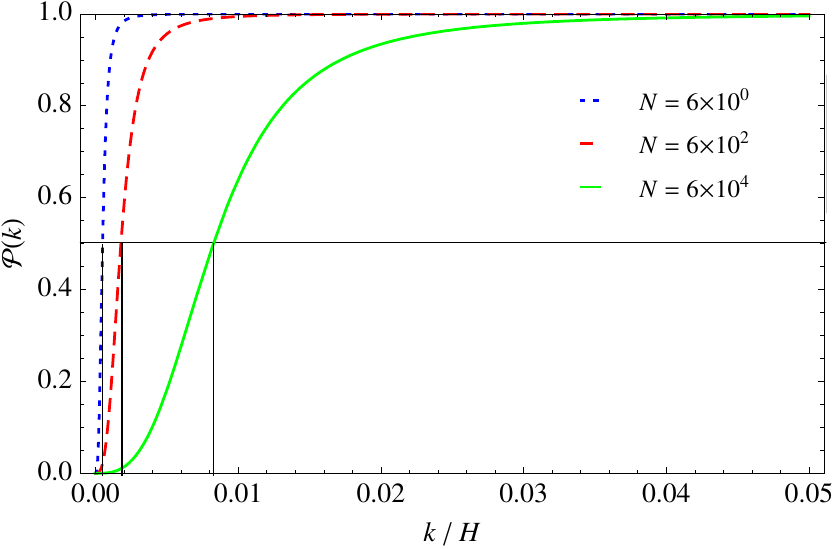}
	\caption{Power spectrum $\Pcal( k )$ of a massless test field with quartic self-coupling as a function of comoving momentum. 
			{\it Upper panel}: Results for various values of $\lambda$, with $N = 6$. 
			{\it Lower panel}: The same as above but for various numbers of $\ee$-folds, where the coupling has been fixed to $\lambda = 10^{-12}$. 
			Also indicated is the value $k^{}_{*}$ at which the power spectrum has decreased to half of its amplitude (normalized to $1$ here).}
	\label{fig:Power-spectrum-lambda-dependence}
\end{figure}

\noindent Its explicit form is rather lengthy and shall not be given here, but the corresponding power spectrum is plotted in Figure \ref{fig:Power-spectrum-lambda-dependence}. At late times one can identify two regimes:
\begin{align}
	\Pcal( t, k )
		&\simeq					\dfrac{ H^{2} }{ ( 2 \pi )^{2} }
								\delimiterfactor=1001
								\begin{cases}
									\vphantom{1^{^{^{^{^{^{^{^{}}}}}}}}}
									\left( \dfrac{ k }{ k^{}_{*}( t ) } \right)^{\!3}	& : \. k \ll k^{}_{*}( t ) ,\\[3.5mm]
									\mspace{33mu} 1 					& : \. k^{}_{*}( t ) \ll k \ll 1 / {\left| \tau( t ) \right|} .
								\end{cases}
								\label{eq:Pcal-late-times}
\end{align}
The comoving wave number $k^{}_{*}$ at which the large-scale behavior of $\Pcal( t, k )$ changes significantly is determined by $\s$ from \eqref{eq:sc=-Delta+4lambdaGaa}. As a function of the number $N$ of quasi-de Sitter $\ee$-folds, for $N$ at least a few and with $N( t = 0 ) \overset{!}{=} 0$, $k^{}_{*}$ reads
\vs{-3mm}
\begin{align}
	k^{}_{*}
		&\simeq					\sqrt[3\mspace{5mu}]{\frac{ \s\mspace{1mu} H }{ 2 }\.}
		\simeq					\sqrt[3\mspace{4.4mu}]{\frac{ 3 }{ 2\.\pi^{2} }\.}\.
								\sqrt[3\mspace{5.5mu}]{\lambda\.N\.}\.H
								.
								\label{eq:khalf}
\end{align}

The factor in front of the curly brace in equation \eqref{eq:Pcal-late-times} is the standard value of the scale-invariant power spectrum. The large-scale behavior [$k \ll k^{}_{*}( t )$] of $\Pcal( t, k )$ follows from the fact that: {a)} the quantity $\s$ is $k$-independent [\cf~\eqref{eq:sc=-Delta+4lambdaGaa}], {b)} the second term in \eqref{eq:Gab} is subdominant compared to the first [hence $\Grm_{ab}( t, k \ra 0 ) = {\rm const.}$], and {c)} the relation of $\Pcal( t, k )$ to $\Grm( t, k )$ involves a factor $k^{3}$.

We observe in Figure \ref{fig:Power-spectrum-lambda-dependence} that the power spectrum is heavily suppressed on large scales in agreement with \eqref{eq:Pcal-late-times}. This damping becomes more pronounced as the self-coupling $\lambda$ is increased and for a large number of $\ee$-folds [\cf~\eqref{eq:khalf}]. Hence, the self-coupling breaks the scale invariance of $\Pcal( t, k )$.

On subhorizon scales one finds $\delimiterfactor=1001 \Pcal\!\left( k \gg 1/ | \tau | \right) \sim k^{2}$. This might be understood from: {a)} the fact that quantum effects in stochastic inflation only significantly modify large scales, {b)} the behavior of the free mode function $\delimiterfactor=1001 u^{}_{0}\!\left( k \gg 1/ | \tau | \right) \sim k^{- 1/2}$, and {c)} the factor $k^{3}$ in \eqref{eq:power-spectrum}.

The derived large-scale suppression solves the problem of infra-red divergencies of real-space correlation functions: While in a scale-invariant theory the two-point function diverges in real space,
\begin{align}
	\Grm( t, \xbm )
		&=						\int\! \frac{ \d^{3}k }{ ( 2 \pi )^{3} }\;
								\ee^{\irm \kbm \cdot \xbm}\.\Grm( t, \kbm )
		\propto					\int \frac{ \d k }{ k }\;\frac{ \sin( k ) }{ k }
		\ra						\infty
								,
								\label{eq:Grm(x)-diverges}
\end{align}
the theory with correctly-resummed quantum effects will be finite. For a complete understanding of quantum effects in inflationary cosmology one would need to include metric fluctuations.

Let us now study if a cut-off at $k^{}_{*}$ could be observable. We assume quasi-de Sitter inflation and a sudden reheating to the radiation-dominated Universe after $N$ $\ee$-folds. This gives for the physical damping scale today,
\begin{align}
	k^{\text{\tiny ph}}_{*}\Big|_{\t{\tiny today}}
		&\simeq					\sqrt[3\mspace{4.4mu}]{\frac{ 3 }{ 2\.\pi^{2} }\.}\.
								\sqrt[3\mspace{5.5mu}]{\lambda\.N\.}\.\ee^{-N}
								\left( \frac{ H }{ T_{\t{\tiny reh}} } \right)\!
								\left( \frac{ T_{0} }{ H_{0} } \right) H_{0}
								.
								\label{eq:khalf-today}
\end{align}
$H_{0}$ is the present value of the Hubble rate, and $T_{\t{\tiny reh}}$ and $T_{0}$ denote the temperatures at reheating and today, respectively. With $\lambda \simeq 10^{-13}$ and $T_{\t{\tiny reh}} \simeq 10^{-3}\.H$ we find for $N \approx 51$ that today the physical damping scale is of the order of the horizon:
\begin{align}
	k^{\text{\tiny ph}}_{*}\Big|_{\t{\tiny today}}
		&\approx					H_{0}
								.
								\label{eq:khalf-today-evaluated}
\end{align}
For a much larger number of $\ee$-folds this cut-off is unobservable.

Other scenarios \cite{Contaldi:2003zv, Ramirez:2009zs} with a finite number of $\ee$-folds also lead to a cut-off in $\Pcal( k )$. However here it is mainly the self-interaction, which is responsible for the large-scale damping. This can be easily seen by considering that the $\l = 0$ result is in our case \eqref{eq:power-spectrum-free-case}.

Suppression of the power spectrum in the infra-red could also influence the cosmic microwave background radiation. This issue was brought into focus by recent observations \cite{Spergel:2003cb, Copi:2008hw, Bennett:2010jb}, which suggest a lack of power on the largest observable scales. Despite a cut-off in the primordial power-spectrum about the Hubble scale, the integrated Sachs-Wolfe effect \cite{Sachs:1967} can regenerate power on the largest observable scales in the cosmic microwave background. Mortonson and Hu \cite{Mortonson:2009xk} recently provided new upper bounds on such a cut-off. They found $k_{\t{\tiny cut}} < 5.2 \times 10^{-4}\.{\rm Mpc}^{-1} $(95\.\% C.L.) using polarization data. This value is close to and well consistent with the above  estimate of $k^{\text{\tiny ph}}_{*} \simeq H_{0} \approx 2.4 \times 10^{-4}\.{\rm Mpc^{-1}}$.

So far we considered tiny self-couplings. However, thanks to the non-perturbative nature of the used methods, our result also applies for stronger couplings. One important example is that of the Higgs boson, which has $\l = \Ocal( 0.1 )$ and{\.---\.}in the perturbative approach{\.---\.}also suffers from infra-red divergences in quasi-de Sitter spaces. In that case $k^{}_{*}$ is much closer to the horizon scale during inflation.

The occurrence of infra-red divergencies in standard perturbation theory is linked to a breakdown of semi-classical methods of quantum field theory in de Sitter space, which has been recently pointed in reference \cite{Burgess:2010dd}. This is shown therein for a scalar field with quartic self-interaction. It has been found that this break-down cannot be cured by including any finite number of loops. It rather concerns the failure of the entire semi-classical calculation.

To summarize, quantum effects in inflationary cosmology significantly modify the large-scale evolution of quantum fields. Within the framework of stochastic inflation and using the methods of replica field theory, we have shown for the specific example of a self-interacting scalar field in de Sitter space-time, that the power spectrum is free from infra-red divergencies due to a large-scale cut-off. Our results are free from any ambiguity associated with the choice of a particular filter function. Furthermore, our findings have been obtained non-perturbatively and are hence not plagued by a break-down of standard perturbation theory in de Sitter space-time (\cf~\cite{Burgess:2010dd}).


It is a pleasure to thank Christian Byrnes, N{\'a}n L{\v \i}, J{\'e}r{\^o}me Martin, Aravind Natarajan, Erandy Ram{\'i}rez and Aleksi Vuorinen for stimulating discussions and support. Furthermore we are very grateful to the anonymous referees for clarifying and enriching comments. We acknowledge support from the DFG and the Alexander von Humboldt foundation.
\vs{-7mm}

\bibliographystyle{apsrev}

\end{document}